# EMPLOYEES ADOPTION OF E-PROCUREMENT SYSTEM: AN EMPIRICAL STUDY

Inder Singh[1] and Devendra Kumar Punia[2]

[1]Center for Information Technology in CES, University of Petroleum & Energy Studies, Dehradun (Uttarakhand), India.
inderddn@yahoo.com

[2]Department of Information Systems in CMES, University of Petroleum & Energy Studies, Dehradun (Uttarakhand), India.

## ABSTRACT

*Today, organizations are investing a lot in their IT infrastructure and reengineering their business processes by digitizing firms. If organizational employees will not optimum utilize its IT infrastructure, the productivity gain reduced enormously. In Uttarakhand e-procurement system implemented by public sector under e-governance integrated mission mode projects. So, there is need to find the determinants which influence employee's adoption and uses of e-procurement systems. This research study assesses the organizational and individual determinants that influence the use of e-procurement system in Uttarakhand public sector. This study provides managers with the valuable information to take intervention programs to achieve greater acceptance and usage of e-procurement system. Data collected for this study by the means of a survey conducted in Uttarakhand state in 2011. A total 1200 questionnaire forms were distributed personally and online to employees using e-procurement system in Uttarakhand.*

## 1. INTRODUCTION

### 1.1 Background

Today, Organizations are investing huge amount in Information Technology (IT), researchers and academicians have struggled to document the organizational gains from IT [17]. Asia-Pacific SMBs spending was $153 billion in 2009 on IT and telecom; More than 50% of Asia-Pacific spending is done by Chinese, Korean, and Indian Small and Medium Businesses [16]. Senior executives have traditionally viewed IT as a back office function that is a "necessary cost" of doing business, without any strategic implications. In the recent years senior managers are now looking IT as a strategic resource and key enabler of growth. Once IT Application tools and software are implemented in the organization, the productivity can only be achieved by acceptance and uses of these tools and software by employees [17].

### 1.2 E-procurement in India

**The vision of Nation e-Governance Plan (NeGP)**, according to [10], "Make all Government services accessible to the common man in his locality, through Common Service Delivery Outlets and ensure efficiency transparency & reliability of such services at affordable costs to realise the basic needs of the common man". Apex Committee Meeting chaired by the Cabinet Secretary reviewed the status and progress of the e-Procurement mission mode projects (MMP), held on 29[th] June 2010. If was the necessity to speed up the implementation of the e-procurement MMP through rigid intervention in the form of directions from the Ministry of Finance by laying down threshold and timeliness for mandatory adoption of e-procurement system for government procurements [10]. NeGP is working on three tiers architecture in which, **Common Service Centres (CSCs)** are the front end delivery point, the second tier provides common and support infrastructure in, including in it are, **Sate Wide Area Networks (SWANs)** facilitate backbone network for data, voice and video and **State Data Centres (SDCs)** provides secure IT infrastructure to host sate state level e-Government application and data. The third tier comprises of 27 Mission Mode Projects (MMPs). Out of this 27 mission mode projects e-procurement MMP is come under integrated MMPs. The vision of the e-Procurement MMP is "To create a national initiative to implement procurement reforms, through the use of electronic Government procurement, so as to make public procurement in all Sector more transparent and efficient" [10].

### 1.3 Benefits of adopting e-procurement

#### 1.3.1 Benefits

Numerous studies proven the potential of e-procurement, according to these researches, "e-procurement facilitate organizations to decentralize their operational procurement processes and centralize strategic procurement processes as a result to provide higher supply chain transparency using e-procurement system" [22]. "Ariba and CommerceOne were founded on the premise that e-procurement software that automates the requisitioning process will be able to reduce processing cost per order from as high as $150 per order to as low as $5 per order" [23]. Compare to tradition procurement transaction using e-procurement can reduce cost per transaction by 65% [8]. According to presutti [21], e-procurement used for inter-organization also enhances the benefits of e-procurement within an organization. Companies using e-procurement system reported that they achieve saving up to 42% in purchasing transaction cost allied with less paperwork, which enables transaction processes to less mistake, and more efficient purchasing. Paper-based procurement process implies transaction costs range from $70 to $300 per purchase order. General Electronic saw those cost drops to 30% by using e-procurement system. Figure 1 is showing the efficiencies generated from the adoption of e-procurement technologies [8].

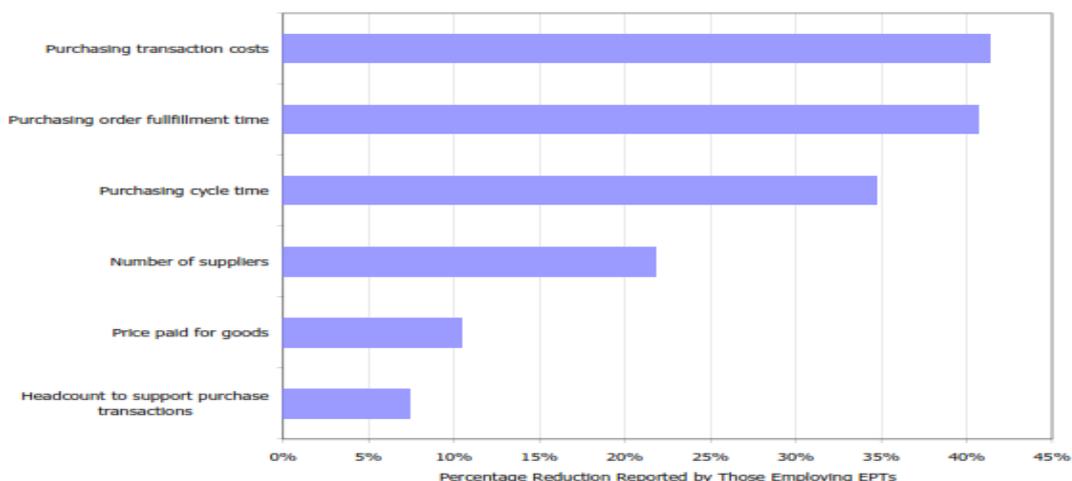

Figure 1. Efficiencies Generated from the Adoption of E-procurement Technologies

Government of Andhra Pradesh achieve many benefits from implementation of e-procurement [3], few of them are, it improves Transparency, Reduced tender cycle time, Saving the Taxpayer's Money, Empowerment of Bidders, Eliminate of Contractors Cartels, Streamlining the Processes, Management Information System.

## 2. LITERATURE REVIEW

### 2.1 The Development of TAM

Researchers are struggling to find out the relationship between IT utilization and its impact on performance of the organizations and individuals [4]. The popular theoretical models in the field of social psychology are Theory of Reasoned Action (TRA) [11] and Theory of Planed Behavior (TPB) [1]. According to TRA [11], an individual's actual behavior is directly influenced by his/her behavioral intention (BI) to use. BI affected by individual's attitude towards that behavior and subjective norm. Attitude defined as "an individual's positive or negative feelings about performing the target behavior" [11]. Whereas, subjective norms defined as "the individual's perception that most people who are important to him think he should not perform the behavior in question". The theory of Planed behavior is an extension to theory of reasoned action which includes another important determinant of behavior, perceived behavioral control [1]. According to TPB [1], perceived behavior control refers to "an individual's perception of the ease or difficulty of performing the behavior of interest ".

There have been three important models widely used by many researchers in IT discipline, to find out the individual's acceptance of IT applications. These three important models are Technology Acceptance Model (TAM) [9], TAM2 – an extension of the Technology Acceptance Model [24], and Unified Theory of Acceptance and Use of technology (UTAUT) [25]. There are two key determinants of TAM [9], Perceived Usefulness (PU) and Perceived Ease of Use (PEOU) which is widely used by the researchers to find the individual's acceptance of IT. According to TAM2 [24], many research has been done to find out the determinants of Perceived Ease of Use, but Perceived Usefulness is a key determinant that is comparatively overlooked. In TAM2, additional theoretical constructs incorporated in TAM that were social influence processes (subjective norm, voluntariness, and image) and cognitive instrumental processes (job relevance, output quality, result demonstrability, and perceived ease of use) which enables organizational managers to organize intervention programs to increase individuals acceptance and usage of new IT applications. According to Venkatesh et. al. [25], eight competing models were reviewed and empirically compared to develop Unified Theory of Acceptance and Use of Technology which identified four constructs that are important determinants of individuals acceptance. Many researchers used these models to find the adoption of new IT Technologies.

### 2.2 E-Procurement

There is rich literature available on e-procurement [7, 15]. By reviewing the literature on e-procurement systems it was found that most of the studies are discussing about the impact of e-procurement systems on organizational performance. But fewer studies have been done on the individual and organizational factors that will affect the adoption of e-procurement system by the employees. The Uttaranchal government's e-governance efforts, the National Informatics Centre (NIC) has in a report highlighted that the project was a "total failure" in major departments. NIC report, based on a study conducted in many departments, found that e-governance was a complete failure [18]. In this study, we discuss about the individual and organizational factors that influence the employee's adoption of e-procurement system in Uttarakhand state.

## 3. RESEARCH METHODOLOGY AND HYPOTHESIS

TAM was extended by many researchers to determine factors affecting the adoption of new IT technologies. In this study, two key determinants of TAM perceived usefulness and perceived ease of use were used to determine the employee's intension to use e-procurement system. In this research study two types of variables were used independent variables and dependable variables. The dependent variable used in this study is behavioral intention to use e-procurement system. The independent variables are used to find the positive or negative impact of these variables on dependent variable. The independent variables used in this research are perceived usefulness, perceived ease of use, infrastructure support, computer self-efficacy, and employee training.

### 3.1 Research Model

The main objective of this study is to determine the factors that will influence the employee's behavioral intention to adopt and use e-procurement system. In this study, two types of factors were used individual and organizational. Individual factor include computer self-efficacy and organizational factors are employees training & infrastructure support was assessed. Two key factors from TAM were also used PU and PEOU. The model is influence by TAM, widely used in technology adoption research. Therefore, TAM determinants are used as a basis for this research model. Once the research model formulated and its relationship identified, hypothesis can be easily propose. Figure 2 clearly explained the relationship between variables.

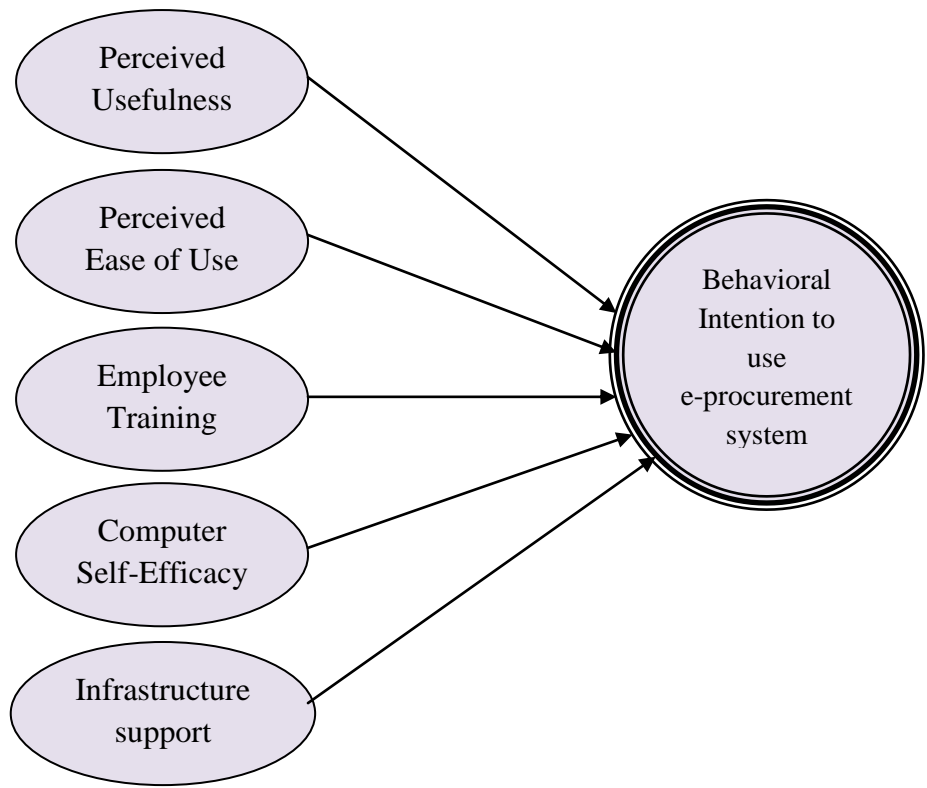

Figure 2. Research Model

## 3.2 Hypothesis

### 3.2.1 Perceived Usefulness

Perceived usefulness refers to "the degree to which a person believes that using a particular system would enhance his or her job performance" [9]. This expectation leads to the following hypothesis:

$H_1$  Perceived Usefulness has a positive effect on behavioral intention to use e-procurement system.

### 3.2.2 Perceived Ease of Use

Perceived ease of use defined as "the degree to which a person believe that using a particular system would be free of effort" [9]. This gives rise to the following the hypothesis:

$H_2$  Perceived Ease of Use has a positive effect on behavioral intention to use e-procurement system.

### 3.2.3 Employee Training

"Training influences user attitudes, behavior, and performance and further that the impact of training on behavioral intention" [12]. Thus, the hypothesis is:

$H_3$  Employee Training has a positive effect on Behavioral intention to use e-procurement system.

### 3.2.4 Computer Self-Efficacy

According to Compeau, D. R., & Higgins, C. A. [5], Computer Self-Efficacy refers to "a judgment of one's capability to use a computer. It is not concerned with what one has done in the past, but rather with judgments of what could be done in the future". This gives rise to the following the hypothesis:

$H_4$  Computer Self-Efficacy has a positive effect on Behavioral intention to use e-procurement system.

### 3.2.5 Infrastructure Support

According to [2], "availability of a well designed infrastructure signals organizational commitment to IT implementation efforts and promotes positive perceptions of IT usefulness and greater satisfaction from IT usage". The hypothesis is:

$H_5$  Infrastructure support has a positive effect on Behavioral intention to use e-procurement system.

## 3.3 Sample

As the population of employees using e-procurement system is large, so that purposeful sampling [26] was used in this study. The study was conducted in Uttarakhand State. The sample taken for this study is from Uttarakhand public sector employees, those who are using e-procurement system. A total of 1200 questionnaire distributed personally and online amongst employees using e-procurement systems tools or applications. Out of 1200 questionnaire 362 questionnaires received back at response rate of 30.17 percent. On further filtering, 345 questionnaires were found completely filled and usable. Each questionnaire item was scored on five-point Likert scale, where (1= Strongly Disagree; 2=Disagree; 3=Neutral; 4=Agree; 5=Strongly Agree).

# 4. DATA ANALYSIS

After checking the completeness and accuracy of data it was fed into Excel sheet, coded, tabulated and analyzed with the help of statistical software Statistical Package for Social Science (SPSS) ver. 18. *Table 1* shows the demographic profiles of the sample.

## 4.1 Demographic Profile and Reliability Analysis

The Reliability of measurement scales was determined by analyses of internal consistency and Cronbach's coefficient alpha (α) test [6]. According to [19], for a measure to be acceptable, the threshold value of coefficient alpha 0.70 or above is sufficient. Table 2 shows the reliability of all the determinants which were found exceeding the minimum threshold value.

Table 1. Demographic Profile of Sample

| Variables | N= 345 | Percent |
|---|---|---|
| **Gender** | | |
| Male | 287 | 83.2 |
| Female | 58 | 16.8 |
| **Age Group** | | |
| Up to 25 years | 33 | 9.6 |
| 26 to 40 years | 130 | 37.7 |
| 41 to 55 years | 158 | 45.8 |
| Above 56 years | 24 | 7.0 |
| **Education Qualification** | | |
| Up to 10+2 | 22 | 6.4 |
| Graduate | 137 | 39.7 |
| Post Graduate | 172 | 49.9 |
| Any Other | 14 | 4.1 |
| **IT Professional** | | |
| Yes | 85 | 24.6 |
| No | 260 | 75.4 |
| **Department** | | |
| Technical | 157 | 45.5 |
| Finance | 66 | 19.1 |
| HR | 49 | 14.2 |
| Others | 73 | 21.2 |
| **Level in the Organization** | | |
| Supervisor | 54 | 15.7 |
| Middle Management | 143 | 41.4 |
| Executive Management | 93 | 27.0 |
| Others | 55 | 15.9 |
| **Number of year using computer** | | |
| 1 to 2 Years | 12 | 3.5 |
| 3 to 5 Years | 52 | 15.1 |
| 6 to 10 Years | 101 | 29.3 |
| More than 10 Years | 180 | 52.2 |
| **Work Experience** | | |
| 1 to 2 Years | 22 | 6.4 |
| 3 to 5 Years | 57 | 16.5 |
| 6 to 10 Years | 77 | 22.3 |
| More than 10 Years | 189 | 54.8 |

Table 2. Reliability Analysis

| Determinants | No. of Items | Sample Reliability |
|---|---|---|
| PU | 04 | .8622 |
| PEOU | 04 | .7710 |
| IS | 03 | .7850 |
| ET | 04 | .7757 |
| CSE | 06 | .7655 |

## 4.2 Factor Analysis

Before proceeding to further factor analysis, appropriateness of factor analysis must be checked. The assessment of factor analysis was done by examining sampling adequacy through Kaiser-Meyer-Olkin (KMO) statistic. According to [14], KMO value greater than 0.60 can be considered as adequate for applying factor analysis. Table 3 shows KMO and Bartlett's test result.

Table 3. KMO and Bartlett's Test Result

| Kaiser-Meyer-Olkin Measure of Sampling Adequacy. | | .888 |
|---|---|---|
| Bartlett's Test of Sphericity | Approx. Chi-Square | 2836.857 |
| | df | 210 |
| | Sig. | 0.000 |

The following values of Bartlett's test result show that the values are significant and thus acceptable. For applying factor extraction, Principle Component Analysis with Varimax rotation and Kaiser Normalization used, factors having Eigen value greater than one were retained.

Table 4 shows the Factor loading, all factors loading are greater than the recommended threshold value 0.45 [13].

Table 4. Factor loading after Varimax rotation with Kaiser Normalization

| Variables | Factors | | | | |
|---|---|---|---|---|---|
| | PU | CSE | ET | IS | PEOU |
| Perceived Usefulness 1 | **.854** | .007 | .101 | .105 | .115 |
| Perceived Usefulness 2 | **.825** | .104 | .120 | .128 | .148 |
| Perceived Usefulness 3 | **.735** | .129 | .151 | .115 | .217 |
| Perceived Usefulness 4 | **.708** | .057 | .269 | .211 | .161 |
| Perceived Ease of Use 1 | .458 | .062 | .185 | .253 | **.514** |
| Perceived Ease of Use 2 | .355 | .101 | .187 | .171 | **.670** |
| Perceived Ease of Use 3 | .064 | .220 | .092 | .116 | **.763** |
| Perceived Ease of Use 4 | .288 | .142 | .217 | .180 | **.652** |
| Infrastructure Support 1 | .271 | -.048 | -.002 | **.756** | .091 |
| Infrastructure Support 2 | .163 | .075 | .174 | **.812** | .118 |
| **Infrastructure Support 3** | .068 | .068 | .148 | **.804** | .256 |
| **Employee Training 1** | .276 | .160 | **.595** | .179 | .124 |
| **Employee Training 2** | .347 | .121 | **.726** | .059 | .068 |
| **Employee Training 3** | .025 | .219 | **.709** | .109 | .140 |
| **Employee Training 4** | .132 | .146 | **.772** | .077 | .238 |
| **Computer Self-Efficacy 1** | .029 | **.742** | -.031 | .045 | .204 |
| **Computer Self-Efficacy 2** | -.018 | **.665** | .096 | .089 | .129 |
| **Computer Self-Efficacy 3** | .219 | **.732** | .077 | .108 | -.018 |
| **Computer Self-Efficacy 4** | -.042 | **.600** | .174 | .035 | .252 |
| **Computer Self-Efficacy 5** | .199 | **.642** | .201 | -.063 | .051 |
| **Computer Self-Efficacy 6** | -.018 | **.567** | .356 | -.133 | -.098 |
| **Eigen Value** | 6.606 | 2.525 | 1.522 | 1.295 | 1.028 |
| **% of variance** | 31.458 | 12.023 | 7.249 | 6.169 | 4.896 |
| **Cumulative %** | 31.458 | 43.481 | 50.729 | 56.898 | 61.794 |
| **Extraction Method: Principal Component Analysis.** <br> **Rotation Method: Varimax with Kaiser Normalization.** <br> **Rotation converged in 6 iterations** | | | | | |

### 4.3 Regression Analysis

The regression analysis was conducted to find, how different factors affect the behavioral intention of employees to use e-procurement system. For this according to [20], the respondent's overall score on behavioral intention to use e-procurement system is considered as dependent variable and other determinants that influence the adoption treated as independent variables. Thus average scores of the different determinants were regressed on the overall score of behavioral intention to use. The beta coefficient provided the relative importance of the determinants. The highest beta coefficient value of a factor considered to have maximum influence on behavioral intention to use e-procurement system while the second highest beta coefficient stands second in terms of relative significance and so on. The overall model was also statistically significant, where ($R^2$ = .405, p<.001).we got the adjusted R Square value 0.397, which shows that this model has accounted for 39.7 % of the variance in the dependent variable. The Regression results are shown in Table 5 and 6.

Table 5. Model Summary

| Model | R | R Square | Adjusted R Square | Std. Error of the Estimate | Change Statistics | | | | |
|---|---|---|---|---|---|---|---|---|---|
| | | | | | R Square Change | F Change | df1 | df2 | Sig. F Change |
| 1 | .637 | .405 | .397 | .47973 | .405 | 46.245 | 5 | 339 | .000 |

a. Predictors: (onstant), CSE, IS, PU, ET, PEOU
b. Dependent Variable: Behavioral Intention to use e-procurement system

Table 6. Coefficients

| Model | | Unstandardized Coefficients | | Standardized Coefficients | t | Sig. |
|---|---|---|---|---|---|---|
| | | B | Std. Error | Beta | | |
| 1 | (Constant) | .524 | .261 | | 2.011 | .045 |

| | | | | | |
|---|---|---|---|---|---|
| PU | .184 | .055 | .180 | 3.372 | .001 |
| PEOU | .361 | .059 | .350 | 6.102 | .000 |
| IS | .125 | .051 | .119 | 2.434 | .015 |
| ET | .114 | .055 | .109 | 2.087 | .038 |
| CSE | .068 | .053 | .061 | 1.279 | .202 |

(a) Dependent Variable: BI

For further testing of our hypothesis correlation analysis was done. Table 7 is showing the correlation analysis values with its level of significance.

Table 7. Correlation Analysis

| Determinants | r | p |
|---|---|---|
| **PU** | .491 | .000 |
| **PEOU** | .584 | .000 |
| **IS** | .404 | .000 |
| **ET** | .427 | .000 |
| **CSE** | .304 | .000 |

The results indicate that Perceived Usefulness, Perceived Ease of Use, Infrastructure Support, Employee Training, and Computer Self-Efficacy have a positive correlation with behavioral intention to use e-procurement system (significance level <0.001).

## 5. DISCUSSION AND FINDINGS

As expected, the result shows that all hypotheses are supported. PU, PEOU, IS, ET, CSE has a positive effect on behavioral intention to use e-procurement system. The result obtained from regression analysis conducted between dependent and independent variables indicate that all the 5 independent factors (PU, PEOU, IS, ET, CSE) have found most influential factors, explaining the intention to use e-procurement system.

When e-procurement system is being find useful which will improves once job performance, the employee's intention to adopt it will be increased. If the e-procurement system is free of effort or easy to use employees are likely to used e-procurement system. The result shows that organization managers should organize some training programs to improve the computer self-efficacy.

## 6. CONCLUSION

The primary objective this research to identify the determinants that influence the behavioral intention of employees to use e-procurement system. TAM is found to be most useful in the field of information systems research for identifying the adoption of new technologies. This study was conducted to explore the factors influencing intention to use e-procurement system. In this research, two key determinants were used from TAM, perceived usefulness and perceived ease of use which have positive effect on intension to use e-procurement system. In order to improve adoption level of employees, public sector managers should develop some intervention programs to improve computer self-efficacy levels of the employees. They should also organize training programs and deploy advance IT infrastructure to support job performance and improve the quality of the employees' work.

## Authors

**Inder Singh**, *M.Tech(IT), M.Sc.(IT), Microsoft certified professional, IBM DB2 certified, e-commerce certification from Asset International.* He is Assistant Professor at University of Petroleum & Energy Studies, Dehradun, India. He has over 10 years of working experience in the area of configuring & troubleshooting computer networks and teaching. His current areas of interest are computer networks, Technology Acceptance, ERP, and e-procurement system.

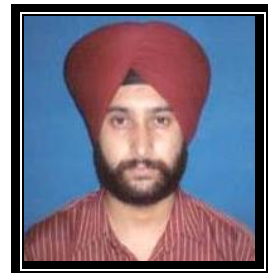

**Dr. Devendra Kumar Punia**, *Bachelor of Engineering in Electronics and Telecommunications from MNIT, Jaipur. He has done his Doctorate in Information Management from Management Development Institute, Gurgaon.* He is Professor and Head of Department of Information Systems in CMES, UPES. He is responsible for academic planning and monitoring, faculty mentoring, research, consultancy and MDPs, budgeting etc. He is a member of Faculty Research Committee and guiding six doctoral students. He is also member of the Joint Coordination Committee for UPES-IBM alliance.

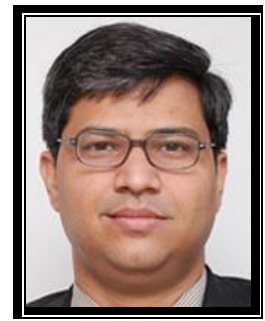